\documentclass[11pt,superscriptaddress,aps,prd,preprint,showpacs,nofootinbib]{revtex4}
\usepackage[dvips]{graphicx}
\usepackage{amsfonts}
\usepackage{slashed}

\usepackage[T1]{fontenc}
\usepackage{ae}
\usepackage[ansinew]{inputenc}
\usepackage{amsmath}
\usepackage{braket}
\usepackage{mathtools}
\usepackage{slashed}
\usepackage{empheq}
\usepackage{tikz}
\usepackage{multirow}
\usetikzlibrary{decorations.pathmorphing}
\usepackage[caption = false]{subfig}

\def\Ss{S\!\!\!\!/}
\def\ds{\partial\!\!\!/}

\newcommand{\be}{\begin{equation}}
\newcommand{\en}{\end{equation}}
\newcommand{\ba}{\begin{eqnarray}}
\newcommand{\ea}{\end{eqnarray}}
\newcommand{\bea}{\begin{eqnarray}}
\newcommand{\eea}{\end{eqnarray}}

\newcommand{\RM}[1]{\mathrm{#1}}

\newcommand{\SLASH}[1]{{#1}\!\!\!/}

\begin{document}

\title{Induced gravitational topological term and the Einstein-Cartan modified theory}

\author{J. R. Nascimento}
\email[]{jroberto@fisica.ufpb.br}
\affiliation{Departamento de F\'{\i}sica, Universidade Federal da 
Para\'{\i}ba,\\
 Caixa Postal 5008, 58051-970, Jo\~ao Pessoa, Para\'{\i}ba, Brazil}

\author{A. Yu. Petrov}
\email[]{petrov@fisica.ufpb.br}
\affiliation{Departamento de F\'{\i}sica, Universidade Federal da Para\'{\i}ba,\\
 Caixa Postal 5008, 58051-970, Jo\~ao Pessoa, Para\'{\i}ba, Brazil}

\author{P. J. Porf\'{\i}rio}
\email[]{pporfirio@fisica.ufpb.br}
\affiliation{Departamento de F\'{\i}sica, Universidade Federal da 
Para\'{\i}ba,\\
 Caixa Postal 5008, 58051-970, Jo\~ao Pessoa, Para\'{\i}ba, Brazil}


\begin{abstract}
It is well known that only the axial piece of the torsion couples minimally to fermions in a Riemann-Cartan geometry, while the other ones decouple. In this paper, we consider the Dirac field minimally coupled to a dynamical background with torsion and compute its contribution to the fermionic one-loop effective action.  Such a contribution owns topological nature since it can be linked with topological invariants from Riemann-Cartan spaces, like Nieh-Yan and Pontryagin (Chern-Pontryagin) terms. Furthermore, we propose a novel modified theory of gravity constructed by adding the aforementioned one-loop contribution to the Einstein-Cartan action. The modified field equations reduce to those ones of GR under certain circumstances, providing therefore trivial solutions. However, in particular, we find a non-trivial solution where the modified field equations do not reduce to the GR ones.     

\end{abstract}


\maketitle

\section{Introduction}
\label{sec:intro}

The technology progress, mainly over the last three decades, allowed important experimental breakthroughs in gravitational physics, as examples one can cite the late-time accelerated expansion of the Universe \cite{Riess}, detection of gravitational waves \cite{Abbott} and direct observations of the shadow of a black hole \cite{Akiyama}. These evidences in some extent corroborate further the astonishing success of the General Relativity (GR) along the years. However, other issues, as for example, the conjecture of the existence of dark matter and dark energy filling the Universe, suggest that GR breaks down in cosmological scales. In this sense, it is believed that alternative theories of gravity, which recover GR in an appropriate limit, could be a promising way for tackling these issues at the cosmic level. In this spirit, a flurry of alternative theories of gravity has been proposed \cite{Albert} from different perspectives, ranging from including new dynamical fields interacting with gravity like \cite{BD} to considering theories defined on non-Riemannian geometries, the so-called metric-affine theories of gravity \cite{Palatini} in which the metric and connection are taken to be independent \textit{a priori}.

One of the first metric-affine model proposed in the literature was the well-known Einstein-Cartan (EC) theory \cite{Hehl:1976kj} where the connection is assumed to possess a non-trivial anti-symmetric counterpart called torsion, in addition to the standard symmetric counterpart (Christoffel symbols) entirely described by the metric. The torsion within this theory is non-dynamical which means that the degrees of freedom associated to it cannot propagate and then their net effects just result in spin-spin contact interactions \cite{Hehl, Hehl2}. In particular, the field equations reduce to the same of GR in the absence of sources. Nonetheless, there be torsion based theories -- teleparallel gravity theories, see \cite{Bahamonde:2021gfp} for a review, where the torsion is permitted to propagate; in addition, more involved theories with dynamical torsion in the Riemann-Cartan geometry have been proposed recently \cite{Katanaev:2020xdv, Bahamonde:2020fnq}. Amongst various motivations to deem torsion as a pivotal ingredient of modified theories of gravity, one can remark the search for a consistent manner of breaking Lorentz-CPT symmetries within gravity context. It has been consistently carried out in the context of the Standard-Model Extension (SME) \cite{KosGra}, where Lorentz-breaking terms involving torsion were proposed in a Riemann-Cartan geometry. 

Topological invariants have historically drawn a great attention in the literature for many reasons \cite{Zanelli:2012px}. Recently, modified theories of gravity based on topological invariants interacting with other fields were proposed. For example, the Pontryagin topological invariant plays a key role in the Chern-Simons modified gravity (CSMG) (see \cite{Jackiw, Yunes} for details), and the Gauss-Bonnet topological invariant, contributing to Einstein-dilaton-Gauss-Bonnet action in the metric formalism (see f.e. \cite{Schiappa} and references therein), have also been formulated for non-trivial torsion \cite{Cambiaso:2010un,Toloza:2012sn,Toloza:2013wi}. It is worth mentioning that these invariants allow to break CPT, and in certain cases Lorentz symmetry \cite{Jackiw}. In a scenario involving non-trivial torsion, it is possible to construct another topological invariant which is the Nieh-Yan term \cite{Nieh, Bombacigno:2021bpk}. Both topological invariants, Pontryagin and Nieh-Yan terms, will be studied in the present paper. We discuss its important aspects, namely, perturbative generation by radiative fermion loops and some exact solution in the EC modified gravity theory involving this invariant as a contact term coupled with the CS topological current.

The structure of the paper looks like follows. In the section 2, we describe fermions minimally coupled to gravity in the Riemann-Cartan geometry. In the section 3, we perform perturbative generation of the Nieh-Yan term, and in the section 4, we obtain classical equations of motion in the modified theory given by a sum of EC and a contact term involving their respective Nieh-Yan and Pontryagin topological currents, and  demonstrate explicitly that the generic spherically symmetric metric solves these equations, as well as G\"{o}del metric is a non-trivial solution of this modified theory. Our conclusions are presented in the section 5.

%
\section{Fermions in a Riemann-Cartan space}
\label{2}

In this section we intend to study  action of a spin-$\frac{1}{2}$ field minimally coupled to gravity with torsion. Unlike the standard approach, where the space-time geometry is taken to be described by a (pseudo)-Riemannian manifold, we will consider the Dirac action defined in a Riemann-Cartan space, i.e., the metric $g_{\mu\nu}$ and the torsion $T^{\alpha}_{\,\,\beta\gamma}$ are treated as independent geometric quantities, see \cite{Hehl:1976kj} for a detailed discussion of the Riemann-Cartan spaces.

Now, we provide the most relevant geometrical tools in Riemann-Cartan spaces. First, let us begin exhibiting the Dirac action minimally coupled to gravity and torsion in a Riemann-Cartan background \cite{Weyl:1950xa, Chandia:1997hu, Shapiro:2001rz}
\begin{equation}
S_{D}=\int d^{4}x\,\sqrt{-g}\left[\frac{i}{2}e^{\mu}_{\,\,a}\left(\bar{\Psi}\gamma^{a}(\nabla_{\mu}^{(\Gamma)}\Psi)-(\nabla_{\mu}^{(\Gamma)}\bar{\Psi})\gamma^{a}\Psi\right)-m\bar{\Psi}\Psi\right],
\label{Dirac}
\end{equation}  
where $\gamma^{a}$ is the usual flat-space Dirac matrices and $e^{a}_{\,\,\mu}$ is the vierbein field. We are choosing the following convention: Latin letters label $SO(1,3)$ group indices running from $0$ to $3$ and Greek letters label space-time indices running from $0$ to $3$. In addition, the metric of the spacetime can be locally defined in terms of a set of orthonormal bases: $\left\{e_{a}(x) \right\}$ and $\left\{\theta^{a}(x) \right\}$,
\begin{equation}
g=\eta_{ab}e^{a}\otimes e^{b}=g_{\mu\nu}dx^{\mu}\otimes dx^{\nu},
\end{equation} 
where $\eta_{ab}$ is the Minkowski metric. The duality condition between both frames leads to the relation: $g_{\mu\nu}=e_{\mu}^{\,\,a}(x)e_{\nu}^{\,\,b}(x)\eta_{ab}$. Similarly, the flat-space Dirac matrices are linked with those ones in curved space through the relations: $\gamma_{\mu}=e_{\mu}^{\,\,a}\gamma_a$ and $\gamma^{\mu}=e^{\mu}_{\,\,a}\gamma^{a}$.  The Fock-Ivanenko covariant derivatives acting on spinors in Eq.(\ref{Dirac}) are defined by
\begin{eqnarray}
\nabla_{\mu}^{(\Gamma)}\Psi&=&\partial_{\mu}\Psi+\Gamma_{\mu}\Psi,\\
\nabla_{\mu}^{(\Gamma)}\bar{\Psi}&=&\partial_{\mu}\bar{\Psi}-\bar{\Psi}\Gamma_{\mu},
\end{eqnarray}
with 
\begin{equation}
\Gamma_{\mu}=\frac{i}{4}\omega_{\mu ab}\sigma^{ab},
\label{gamma}
\end{equation}
where $\omega_{\mu ab}$ are the components of the spin connection and $\sigma^{ab}=\frac{i}{2}[\gamma^{a},\gamma^{b}]$ are the generators of the covering Lorentz group in the spinor representation.

In order to proceed further it makes necessary to define other important geometrical quantities. For that, let us write down the Cartan structure equations that summarize the main properties of the Riemann-Cartan geometry,
\begin{eqnarray}
0&=&D\eta_{ab};\\
T^{a}&=&D\theta^{a}=d\theta^{a}+\omega^{a}_{b}\wedge \theta^{b};\\
R^{a}_{\,\,b}&=&D\omega^{a}_{\,\,b}=d\omega^{a}_{\,\,b}+\omega^{a}_{\,\,c}\wedge\omega^{c}_{\,\,b},
\end{eqnarray}
 where $\omega^{a}_{\,\,\,b}=\omega_{\mu\,\,\,b}^{\,\,\,a}dx^{\mu}$ is the connection one-form, $T^{a}=\frac{1}{2}T^{a}_{\,\,\mu\nu}dx^{\mu}\wedge dx^{\nu}$ is the  torsion two-form and $R^{a}_{\,\,b}=\frac{1}{2}R^{a}_{\,\,b\mu\nu}dx^{\mu}\wedge dx^{\nu}$ is the curvature two-form. Note that the first Cartan equation is indeed a constraint on the spin connection resulting in $\omega_{\mu ab}=-\omega_{\mu ba}$. The second Cartan equation relates the torsion tensor with the tetrad and the spin connections, in terms of components one can write down
\begin{equation}
T^{a}_{\,\,\mu\nu}=\partial_{\mu}e^{a}_{\,\,\nu}-\partial_{\nu}e^{a}_{\,\,\mu}+\omega_{\mu\,\,\,\nu}^{\,\,\,\,\,a}-\omega_{\nu\,\,\,\mu}^{\,\,\,\,\,a}.
\end{equation}  
 
The spin connection can be decomposed into two parts in Riemann-Cartan geometry,
\begin{equation}
\omega_{\mu}^{\,\,\,ab}=\tilde{\omega}_{\mu}^{\,\,\,ab}+K^{ba}_{\,\,\,\,\mu},
\label{omega}
\end{equation}
   where $\tilde{\omega}_{\mu}^{\,\,\,ab}$ is the torsionless Cartan connection which is entirely determined by the vierbeins. Its explicit form is given by
	\begin{equation}
	\tilde{\omega}_{cab}=-\Omega_{cab}-\Omega_{acb}+\Omega_{bca},
	\end{equation}
where $\Omega_{abc}=\frac{1}{2}e^{\,\,\mu}_{b}e^{\,\,\nu}_{c}\left(\partial_{\mu}e_{\nu a}-\partial_{\nu}e_{\mu a}\right)$ are the anholonomy or Ricci rotation coefficients. The second term on the \textit{r.h.s.} of Eq.(\ref{omega}) is precisely the contorsion tensor,
\begin{equation}
K^{\mu}_{\,\,\alpha\beta}=\frac{1}{2}\left(T^{\mu}_{\,\,\alpha\beta}-T_{\beta\,\,\,\,\alpha}^{\,\,\,\mu}-T_{\alpha\,\,\,\,\beta}^{\,\,\,\mu}\right),
\label{contorsion}
\end{equation}  
in which is antisymmetric in the first two indices, $K_{\mu\nu\alpha}=-K_{\nu\mu\alpha}$ and $K^{\mu}_{\,\,\alpha\beta}-K^{\mu}_{\,\,\beta\alpha}=T^{\mu}_{\,\,\alpha\beta}$.

For our purposes it is convenient to rewrite Eq. (\ref{Dirac}) in a more appropriate way. Keeping this in mind, we integrate it by parts and substitute Eqs. (\ref{omega}) and (\ref{contorsion}) into Eq. (\ref{Dirac}) to find
\begin{equation}
\begin{split}
S_{D}&=\int d^{4}x\,\sqrt{-g}\,\bar{\Psi}\left[i\gamma^{\mu}\tilde{\nabla}_{\mu}+\frac{1}{8}S_{\mu}\gamma_{5}\gamma^{\mu}-m\right]\Psi\\
&=\tilde{S}_{D}+\frac{1}{8}\int d^{4}x\,\sqrt{-g}\,S_{\mu}\bar{\Psi}\gamma_{5}\gamma^{\mu}\Psi\\
&=\tilde{S}_{D}-\frac{1}{8}\int d^{4}x\,\sqrt{-g}\,S_{\mu}J^{\mu}_{5},
\label{Dirac2}
\end{split}
\end{equation}  
where $\tilde{S}_{D}$ stands for the usual Dirac action defined in terms of the Cartan connection as we shall see later and $J^{\mu}_{5}\equiv \bar{\Psi}\gamma^{\mu}\gamma_{5}\Psi$ is the axial spin vector current. By virtue of the above definitions, the Fock-Ivanenko covariant derivatives acting on spinors are now written as follows
\begin{equation}
\begin{split}
\tilde{\nabla}_{\mu}\Psi&=\partial_{\mu}\Psi+\tilde{\Gamma}_{\mu}\Psi,\\
\tilde{\nabla}_{\mu}\bar{\Psi}&=\partial_{\mu}\bar{\Psi}-\bar{\Psi}\tilde{\Gamma}_{\mu},
\end{split}
\end{equation}
with $\tilde{\Gamma}_{\mu}$ defined similar to Eq.(\ref{gamma}) just with $\tilde{\omega}_{\mu}^{\,\,\,ab}$ replacing $\omega_{\mu}^{\,\,\,ab}$. Furthermore, we also defined the axial-vector torsion $S_{\mu}=\epsilon_{\mu\nu\alpha\beta}T^{\nu\alpha\beta}$. Therefore, for fermions minimally coupled to gravity and torsion, only the axial part of the torsion couples to fermions in Riemann-Cartan spaces \cite{Shapiro:2001rz, Obukhov}. 

In the context of the Standard Model Extension (SME) \cite{KosGra}, the 
 spinor-pseudovector interaction term can be typically re-interpreted as a CPT-violating one since this interaction mimics the axial coefficient for Lorentz/CPT violation $b_{\mu}$ \cite{KosGra}. In this sense, the effect of the torsion tensor emerges as an external background field in which can be identified by $b_{\mu}=\frac{1}{8}S_{\mu}$ in the SME. This connection between the CPT-violating coefficient and the background torsion suggests that experiments estimating $b_{\mu}$ could provide information on the nature of the space-time geometry, namely, whether it is metric based (pseudo-Riemannian) -- that corresponds to very stringent estimations for $b_{\mu}$ -- or a Riemann-Cartan geometry. Various experiments have been proposed  in order to estimate the parameter $b_{\mu}$ \cite{Mohanty}.

In the non-relativistic limit such an interaction term mimics a sort of Zeeman effect, then, describing an interaction between the fermion spin with the external background field ($\vec{b}$) \cite{Mukhopadhyay, rub}. In the next section we will address the one-loop corrections to the effective fermionic action treating the axial-vector torsion as an external field.

\section{ONE-LOOP INDUCED GRAVITATIONAL TOPOLOGICAL TERM}

We aim at this section to show the induced gravitational topological term upon integrating the fermions out in the effective action at one-loop level. To start with {the calculation, let us rewrite Eq. (\ref{Dirac2}) as follows:
\begin{equation}
S_{D}=\int d^{4}x\,\sqrt{-g}\,\bar{\Psi}\left[i\slashed{\partial}-\frac{1}{8}\slashed{S}\gamma_{5}+\tilde{\slashed{\Gamma}}-m\right]\Psi,
\end{equation} 
where $\tilde{\Gamma}_{\mu}=-\frac{1}{4}\tilde{\omega}_{\mu ab}\sigma^{ab}$.  The next calculations are performed along the same lines as in the paper \cite{ptime}: we can rewrite this action within the tetrad formalism as
\begin{equation}
	S_{D}=\int d^{4}x\,e e^{\mu}_a\,\bar{\Psi}\left[i\tilde{\nabla}_{\mu}\gamma^a-\frac{1}{8}S_{\mu}\gamma^a\gamma_{5}-m\right]\Psi,
\end{equation} 
which allows us to write the one-loop effective action in the form
\begin{equation}
	\label{trace}
	\Gamma^{(1)}=-i{\rm Tr}\ln(i\tilde{\slashed{\nabla}}-m-\frac{1}{8}\Ss\gamma_5)
\end{equation}
This action is treated with use of the derivative expansion formalism \cite{Das}. Actually it means that the functional trace (\ref{trace}) must be expanded up to the first order in derivatives. Moreover, the zero order can be disregarded since both connection and torsion have odd numbers of indices and product of a torsion and two connections, without derivatives, cannot form a scalar object.

To proceed with this calculation, we follow  a  manner similar to that one employed in \cite{ptime}. First of all, following the standard approach, we must rewrite the one-loop effective action as a trace of logarithm of some second-derivative operator. In this case, however, the situation is slightly different: since the desired term involves both Cartan connection and torsion dependence, so that in the zero torsion case it vanishes, we can add the term which depends explicitly only on the torsion vector, that is,
\begin{equation}
	\Gamma_0[S]=-i{\rm Tr}\ln(i\ds+m+\frac{1}{8}\Ss\gamma_5).
\end{equation}
We note that this term, although it is not a constant, will not affect the desired term since it will depend on $S_{\mu}$ only, and, being a scalar, it can yield only $S^2$ and higher terms like $(\partial S)^2$, $S^2\partial S$, $S^4$, etc (for further details, we refer the reader to \cite{Ojima:1988av, ptime}).
Therefore, the resulting term can be presented as
\bea
\Gamma_{NY}&=&-i{\rm Tr}\ln(i\tilde{\slashed{\nabla}}-m-\frac{1}{8}\Ss\gamma_5)-i{\rm Tr}\ln(i\ds+m+\frac{1}{8}\Ss\gamma_5)|_{NY}=\nonumber\\
&=&-i{\rm Tr}\Big[
-\Box+i\tilde{\slashed{\Gamma}}\ds+m\tilde{\slashed{\Gamma}}-m^2+\frac{1}{8}(\tilde{\slashed{\Gamma}}-2m)\Ss\gamma_5+\frac{2i}{8}(S\cdot\partial)\gamma_5-\frac{1}{64}S^2
\Big].
\label{eqt}
\eea
We shall proceed along the same lines as in \cite{ptime}, i.e., one expands the last expression in power series of $S_{\mu}$ up to the first order, and in the connection up to the second order. In general, this trace diverges, thus, a regularization procedure must be used in order to split the finite and divergent contributions. Here, we shall follow the Schwinger-DeWitt proper time method \cite{Barvinsky:1985an} and also focus only on the finite contribution stemming from Eq. (\ref{eqt}). 

Using the Schwinger-DeWitt method we are able to find that the first finite contribution of Eq.(\ref{eqt}) for the topological Chern-Simons term is
\bea
\Gamma_{NY,2}&=& \RM{tr}_\RM{D} \int d^4x\int_0^{\infty}ds\,e^{-sm^2}\left[ \frac{1}{8}s^2m^2\SLASH{\tilde{\Gamma}}(\SLASH{\partial}\SLASH{\tilde{\Gamma}})\SLASH{S}\gamma_5+\frac{1}{4}m^2s^3\SLASH{\tilde{\Gamma}}\SLASH{\partial}(\partial_\alpha\SLASH{\tilde{\Gamma}})\partial^\alpha\SLASH{S}\gamma_5\right. \nonumber\\
&&\left.+\frac{1}{4}m^2s^3\SLASH{\tilde{\Gamma}}(\partial_\alpha\SLASH{\tilde{\Gamma}})\partial^\alpha\SLASH{\partial}\SLASH{S}\gamma_5\right]e^{-s\Box}\delta(x-x')\big|_{x'=x}.
\label{NY2}
\eea
Following the same definitions used in \cite{heat}, the delta function in curved spaces is defined as below
\bea
\delta(x-x^{\prime})=\int \frac{d^{4}k}{(2\pi)^4} e^{ik_{\mu}\tilde{\nabla}^{\mu}\sigma(x,x^{\prime})}, 
\eea
where $\sigma(x,x^{\prime})$ is the geodesic distance satisfying the relation $\frac{1}{2}\tilde{\nabla}_{\mu}\sigma(x,x^{\prime})\tilde{\nabla}^{\mu}\sigma(x,x^{\prime})=\sigma(x,x^{\prime})$ and defined in such a way that
\bea
\lim_{x{\rightarrow}x^{\prime}}\tilde{\nabla}^{\mu}\tilde{\nabla}^{\nu}\sigma(x,x^{\prime})=g^{\mu\nu}.
\eea
Upon carrying out the trace over the Dirac gamma matrices and integration over the proper time $s$ (the step-by-step calculation can be made similarly to \cite{ptime}), we obtain the first contribution to the desired effective action:
\bea
\Gamma_{NY,2}&=&\frac{1}{216\pi^2}\int d^4x\sqrt{-g}\,\epsilon^{\mu\nu\lambda\rho}S_{\mu}\partial_{\nu}\tilde{\omega}_{\lambda ab}\tilde{\omega}_{\rho}^{ab}.
\label{ny2}
\eea
The second contribution arises} keeping only derivative independent terms, but up to the third order in connection, namely:
\bea
\Gamma_{NY,3}&=&i\,\RM{tr}_\RM{D}\! \int d^4x \int_0^{\infty}ds\,e^{-sm^2}\left[\frac{s^2}{8}m^2\SLASH{\tilde{\Gamma}}\SLASH{\tilde{\Gamma}}\SLASH{\tilde{\Gamma}}\SLASH{S}\gamma^5+\frac{s^3}{24}m^2\left(\SLASH{\tilde{\Gamma}}\SLASH{\tilde{\nabla}}\SLASH{\tilde{\Gamma}}\SLASH{\tilde{\nabla}}\SLASH{\tilde{\Gamma}}+\SLASH{\tilde{\Gamma}}\SLASH{\tilde{\nabla}}\SLASH{\tilde{\Gamma}}\SLASH{\tilde{\Gamma}}\SLASH{\tilde{\nabla}}+\SLASH{\tilde{\Gamma}}\SLASH{\tilde{\Gamma}}\SLASH{\tilde{\nabla}}\SLASH{\tilde{\Gamma}}\SLASH{\tilde{\nabla}}\right)\SLASH{S}\gamma_5\right.\nonumber\\
&&\left.-\frac{s^3}{24}m^2\left(\SLASH{\tilde{\Gamma}}\SLASH{\tilde{\nabla}}\SLASH{\tilde{\Gamma}}\SLASH{\tilde{\Gamma}}+\SLASH{\tilde{\Gamma}}\SLASH{\tilde{\Gamma}}\SLASH{\tilde{\nabla}}\SLASH{\tilde{\Gamma}}+\SLASH{\tilde{\Gamma}}\SLASH{\tilde{\Gamma}}\SLASH{\tilde{\Gamma}}\SLASH{\tilde{\nabla}}\right)S\cdot \tilde{\nabla}\gamma_5-\frac{s^3}{24}m^4\SLASH{\tilde{\Gamma}}\SLASH{\tilde{\Gamma}}\SLASH{\tilde{\Gamma}}\SLASH{S}\gamma^5\right]e^{-s\Box}\delta(x-x')\big|_{x'=x}.
\label{NY3}
\eea
 Proceeding in a similar way to the former contribution we found  
\bea
\Gamma_{NY,3}&=&\frac{1}{324\pi^2}\int d^4x\sqrt{-g}\,\epsilon^{\mu\nu\lambda\rho}S_{\mu}\tilde{\omega}_{\nu ab}\tilde{\omega}_{\lambda}^{bc}\tilde{\omega}_{\rho c}^{\phantom{\rho c}a}.
\label{ny3}
\eea
The sum of the above two finite contributions is given by 
\bea
\Gamma_{NY}&=&\frac{1}{216\pi^2}\int d^4x\sqrt{-g}\,\epsilon^{\mu\nu\lambda\rho}S_{\mu}\Big(\partial_{\nu}\tilde{\omega}_{\lambda ab}\tilde{\omega}_{\rho}^{ab}+\frac{2}{3}
\tilde{\omega}_{\nu ab}\tilde{\omega}_{\lambda}^{bc}\tilde{\omega}_{\rho c}^{\phantom{\rho c}a}\Big).
\label{ny}
\eea
is a contact interaction term. Indeed, (\ref{ny}) represents the interaction between two topological currents, namely, Nieh-Yan and Pontryagin currents. The origin of both lies on their respective topological invariants. As it is known, the Nieh-Yan and Pontryagin topological terms are defined by 
\begin{eqnarray}
\label{Ny}NY&=&T^{a}\wedge T_{a}-e^{a}\wedge e^{b}\wedge R_{ab},\\
P&=&R^{a}_{\,\,b}\wedge R^{b}_{\,\,a},
\end{eqnarray}
respectively. Note that the second term on the \textit{r.h.s} of Eq.(\ref{Ny}) is the well-known Holst term \cite{Shapiro}. Their topological natures are settled by rewriting them as total derivatives:
\begin{eqnarray}
NY&=&dQ, \,\,\, Q=e^{a}\wedge T_{a},\\
P&=&dC, \,\,\, C=\omega^{a}_{\,\,b}\wedge d\omega^{b}_{\,\,a}+\frac{2}{3}\omega^{a}_{\,\,c}\wedge\omega^{c}_{\,\,b}\wedge\omega^{b}_{\,\,a}, 
\end{eqnarray}
where $Q$ and $C$ are the Nieh-Yan and Chern-Simons three-form topological currents. These invariants can be rewritten in terms of components as follows
\begin{eqnarray}
NY&=&-\frac{1}{2}\tilde{\nabla}_{\mu}S^{\mu},\\
P&=&2\tilde{\nabla}_{\mu}C^{\mu},
\end{eqnarray}
where 
\begin{equation}
C^{\mu}=\epsilon^{\mu\nu\lambda\rho}\left[\omega_{\rho}^{\,\,ba}\partial_{\nu}\omega_{\lambda ab}+\frac{2}{3}\omega_{\nu ab}\,\omega_{\lambda}^{\,\,bc}\,\omega_{\rho c}^{\,\,\,\,a}\right]
\label{Cc}
\end{equation}
is the topological Chern-Simons vector current. Similarly, the axial piece of the torsion is the Nieh-Yan topological current. These definitions allows us to rewrite Eq. (\ref{ny}) as an axial-axial contact interaction term coupling both topological currents
\begin{equation}
\Gamma_{NY}=\frac{1}{216\pi^2}\int d^4x\sqrt{-g}\,S_{\mu}\tilde{C}^{\mu},
\label{NY}
\end{equation} 
 where $\tilde{C}^{\mu}$ is defined as in Eq.(\ref{Cc}) just with the torsionless spin connection $\tilde{\omega}_{\mu}^{\,\,ab}$ substituting $\omega_{\mu}^{\,\,ab}$.
Since the axial vector part of the torsion (Nieh-Yan current) is assumed to be a background vector field, then, one can interpret the fermionic one-loop effective action (\ref{NY}) as an interaction between the background space-time torsion -- more precisely, the axial-vector torsion --  with the Chern-Simons current $\tilde{C}^{\mu}$. In this situation, $S_{\mu}$, as it was said before, can be interpreted as a particular coefficient for local Lorentz/CPT violation from SME \cite{KosGra}. The more important physical effect is that any observable (in our case $\tilde{C}_{\mu}$) which couples to $S_{\mu}$ will feel the Lorentz-violating effects through the axial-vector torsion. Thus, in this sense, the contact term is an explicit local Lorentz-violating term. It is not a surprising result since as long as we turn off the background torsion which, as we have remarked before, plays the role of the axial field in \cite{ptime}, this effective interaction vanishes, in much the same way to take the Lorentz-violating coefficient $b_{\mu}=0$ in \cite{ptime}, then our results are in agreement with those found in \cite{ptime} for the pseudo-Riemannian geometry. On the other hand, maintaining a non-zero background torsion, the effective action at the one-loop level  behaves as an axial background field as pointed out before. Accordingly, we conclude that this interaction emerges as a purely geometrical effect in a way different from that one in \cite{ptime}, where the background field $b_{\mu}$ is set by hand in the action.

It is worth to mention that this novel contact term resembles the interaction term proposed within the non-dynamical version of CSMGR \cite{Jackiw, Yunes}, with the axial-vector torsion background field playing the role of the axial vector field $v_{\mu}$ defined in those papers. However, in our case the situation looks like thoroughly different because the background field appears naturally as a result of the modification of the space-time geometry as we have already mentioned, whilst $v_{\mu}$ is just an external quantity fixed \textit{a priori} and then without any relation with the other geometrical quantities. 

\section{EINSTEIN-CARTAN MODIFIED THEORY WITH ADDITIVE TOPOLOGICAL TERM}

In this section we propose a simple modification of the EC theory inspired by a whole analogy with effective field theories \cite{Georgi:1993mps}. Our modification proposal comprises of adding the contact term generated by the one-loop quantum correction (\ref{NY}), obtained in the previous section, to the EC action. To wit,
\begin{equation}
S_{ECM}=\frac{1}{4\kappa^2}\int \left(\epsilon_{abcd} e^{a}\wedge e^{b}\wedge R^{cd}+\alpha S^{a}\wedge (\star \tilde{C})_{a} \right)+S_{sources}+...,
\end{equation}
where $\star$ means the Hodge star operator, $S_{sources}$ is the action of matter and spin sources and the ellipsis stands for subleading terms at low energy level which contain dynamical and higher-order torsion terms. Nonetheless, from the phenomenological point of view, one can suppose that dynamical torsion terms are highly (Planck scale) suppressed since there is no experimental evidence corroborating the propagation of the torsion \cite{Hehl:1976kj}. Consequently, at the classical level, we can safely truncate our effective modified EC theory to the leading terms displaying in the former action explicitly. In the context of effective field theories, the real parameter $\alpha=\frac{1}{\Lambda^2}$, where $\Lambda$ is a typical high energy (UV) scale. Hence, the previous equation (disregarding the subleading terms written in terms of components becomes 
\begin{equation}
\begin{split}
S_{ECM}=\frac{1}{2\kappa^2}\int d^{4}x\sqrt{-g} \left(R(\Gamma)+\frac{\alpha}{12} S_{\mu}\tilde{C}^{\mu}\right) +S_{sources},
\label{con}
\end{split}
\end{equation} 
where $R(\Gamma)=g^{\mu\nu}R_{\mu\nu}(\Gamma)$ is the Ricci scalar of the full affine connection $\Gamma^{\mu}_{\,\,\alpha \beta}$ which, in turn, is related to the Levi-Civita connection ($L^{\mu}_{\,\,\alpha \beta}$) and the contorsion tensor by: $\Gamma^{\mu}_{\,\,\alpha\beta}=L^{\mu}_{\,\,\alpha \beta}+K^{\mu}_{\,\,\alpha \beta}$. Taking these definitions into account, Eq.(\ref{con}) can be written as
 \begin{equation}
S_{ECM}=\frac{1}{2\kappa^2}\int d^{4}x\sqrt{-g}\left( \tilde{R}+2\tilde{\nabla}_{\lambda}K^{\lambda}+K_{\alpha\beta\gamma}K^{\alpha\gamma\beta}-K_{\lambda}K^{\lambda}+\frac{\alpha}{12} S_{\mu}\tilde{C}^{\mu}\right)+ S_{sources},
\label{con1}
\end{equation}
where $\tilde{R}=g^{\mu\nu}R_{\mu\nu}(L)$ is the Ricci scalar related to the Levi-Civita connection and $K_{\lambda}\equiv K^{\tau}_{\,\,\lambda\tau}$. Varying the action (\ref{con1}) with respect to the metric and the contorsion, respectively, we find the field equations for the modified theory 
\begin{eqnarray}
\label{32}\tilde{R}_{\mu\nu}-\frac{1}{2}g_{\mu\nu}\tilde{R}-\frac{\alpha}{24}X_{\mu\nu}&=&k^{2}T_{\mu\nu},\\
\label{33}T^{\alpha}_{\,\,\beta\gamma}+\delta^{\alpha}_{\gamma}T_{\beta}-\delta^{\alpha}_{\beta}T_{\gamma}-\frac{\alpha}{6}\epsilon^{\alpha}_{\,\,\mu\gamma\beta}\tilde{C}^{\mu}&=&\kappa^{2}\Theta^{\alpha}_{\,\,\beta\gamma},
\end{eqnarray}
where 
\begin{equation}
T_{\mu\nu}=-\frac{2}{\sqrt{-g}}\frac{\delta S_{sources}}{\delta g^{\mu\nu}}
\end{equation} 
 is the energy-momentum tensor and
\begin{equation}
\Theta^{\alpha}_{\,\,\beta\gamma}=-\frac{2}{\sqrt{-g}}\frac{\delta S_{sources}}{\delta K_{\alpha}^{\,\,\beta\gamma}}
\end{equation}
 is the spin tensor. Note that from Eq. (\ref{32}) the variation with respect to the Chern-Simons current yields a new symmetric tensor which is defined by
\begin{equation}
X^{\mu\nu}=\left(S_{\lambda}\epsilon^{\lambda\beta\gamma\nu}\tilde{\nabla}_{\beta}\tilde{R}^{\mu}_{\gamma}-(\tilde{\nabla}_{\sigma}S_{\lambda})\,^{*}\tilde{R}^{\sigma\mu\lambda\nu}+(\mu\leftrightarrow\nu)\right),
\end{equation}
where $\,^{*}\tilde{R}^{\sigma\mu\lambda\nu}=\frac{1}{2}\epsilon^{\sigma\mu}_{\,\,\,\,\,\,\,\alpha\beta}R^{\alpha\beta\lambda\nu}$ is the dual Riemann tensor. The tensor $X_{\mu\nu}$ is sometimes called the Cotton tensor \cite{Jackiw}. The Eq. (\ref{33}) provides an algebraic equation for the torsion similarly to the situation occurring in the EC theory, however in our case there is an extra term coming from the contact interaction one.  

To better understand the influence of the role played by the torsion in the modified theory let us turn our attention to the vacuum field equations, namely, in the absence of sources. In this context, the field equations reads
\begin{eqnarray}
\label{37}\tilde{R}_{\mu\nu}-\frac{1}{2}g_{\mu\nu}\tilde{R}-\frac{\alpha}{24}X_{\mu\nu}&=&0,\\
\label{k}T^{\alpha}_{\,\,\beta\gamma}+\delta^{\alpha}_{\gamma}T_{\beta}-\delta^{\alpha}_{\beta}T_{\gamma}-\frac{\alpha}{6}\epsilon^{\alpha}_{\,\,\mu\gamma\beta}\tilde{C}^{\mu}&=&0.
\end{eqnarray}
Upon taking the trace ($\alpha=\beta$) in the second equation we are able to find that $T_{\gamma}=0$. Inserting this in Eq. (\ref{k}) one obtains 
\begin{equation}
T_{\alpha\beta\gamma}=\frac{\alpha}{6}\epsilon_{\alpha\mu\gamma\beta}\tilde{C}^{\mu},
\label{er}
\end{equation}
or by dualizing it
\begin{equation}
S^{\mu}=-\alpha \tilde{C}^{\mu}.
\label{ss}
\end{equation} 
This equation is in fact a constraint equation, as a result, it settles that the axial-vector torsion is completely determined by the Chern-Simons topological current of the spinless connection. Such a result is not a surprise because, formally, the Chern-Simons topological current could be safely interpreted as an external source for the torsion similarly to what happens as we allow matter sources to couple to torsion \cite{Shapiro:2001rz}. Therefore, differently from the EC theory, the vacuum field equations of the modified theory entail in the constraint (\ref{ss}) which enforces the axial-vector torsion to be proportional to the Chern-Simons topological current. Apart from that, the divergence of Eq. (\ref{37}) imposes another constraint or consistency condition to hold the diffeomorphism invariance of the modified theory 
\begin{equation}
\tilde{\nabla}_{\mu}X^{\mu\nu}=-\frac{\alpha}{4}\tilde{C}^{\nu}\tilde{\nabla}_{\mu}\tilde{C}^{\mu}=0,
\end{equation}
 which is somewhat similar to the Pontryagin constraint in CSMGR \cite{Jackiw}. Hence consistent solutions must be restricted to the parameter space 
corresponding to the vanishing of either $\tilde{C}^{\mu}$ or $\tilde{\nabla}_{\mu}\tilde{C}^{\mu}$.

Substituting Eq. (\ref{ss}) into the Eq. (\ref{37}), we are able to eliminate the dependence of the Eq. (\ref{37}) on the axial piece of the torsion arriving at
	\begin{equation}
		\label{eqmot}
		\tilde{R}_{\mu\nu}-\frac{1}{2}g_{\mu\nu}\tilde{R}+\frac{\alpha^2}{24}\left[\tilde{C}_{\lambda}\epsilon^{\lambda\beta\gamma}_{\phantom{\lambda\beta\gamma}\nu}\tilde{\nabla}_{\beta}\tilde{R}_{\mu\gamma}-(\tilde{\nabla}_{\sigma}S_{\lambda})\,^{*}\tilde{R}^{\sigma
			\phantom{\mu}\lambda\phantom{\nu}}_{\phantom{\sigma}\mu\phantom{\lambda}\nu}+(\mu\leftrightarrow\nu) \right]=0.
	\end{equation}

Note that the contact term leads to third-order derivatives of the metric in the field equations, so it can potentially develop ghost degrees of freedom. However, such potential ghosts are very heavy in the regime of validity of the effective field theory, thus, their impact can be neglected at the low-energy limit. Because of that, one can interpret the previous contact interaction as a particular higher-order Lorentz-breaking term \cite{KosGra}. 

To clarify further our analysis let us solve the field equations for a generic spherically symmetric static metric. A straightforward computation shows that in this case all components of $\tilde{C}^{\mu}$ vanish and then one recovers the GR field equations. Such a result is expected since the CS topological current seems to be sensitive only for spinning metrics as, for example, the Kerr one that provides $\tilde{C}^{\mu}\neq 0$, but does not solve the equations of motion for the modified theory due to the fact that the consistency condition fails since $\tilde{\nabla}_{\mu}\tilde{C}^{\mu}\neq 0$. In particular, Schwarzschild metric persists as a solution of our EC modified theory, in contrast to the modified theory proposed in \cite{BottaCantcheff:2008pii}. Another interesting example is the well-known rotating G\"{o}del metric \cite{Godel:1949ga}, in this case, the component $\tilde{C}^{z}\neq 0$, however it satisfies the other consistency condition: $\tilde{\nabla}_{\mu}\tilde{C}^{\mu}=0$. It is worthwhile to remark that the field equations of the modified theory do not reduce to the GR ones because $X_{\mu\nu}\neq 0$. Therefore, even the G\"{o}del metric being a solution for the GR field equations in the presence of well-motivated matter sources and cosmological constant \cite{Godel:1949ga}, it is also a non-trivial solution of our modified theory in the presence of other kinds of matter sources similarly to what happens in CSMGR \cite{ourgodel}.

 \section{SUMMARY AND CONCLUSIONS}\label{6}

We have tackled with quantum and classical aspects of fermions minimally coupled to gravity with torsion. First, we computed the fermionic one-loop effective action by the proper time method in a full analogy to \cite{ptime} and we have found a finite contribution since it is superficially divergent. Remarkably, upon integrating the fermions out, the fermionic one-loop effective action results in a contact interaction term between two topological terms, namely: the axial-vector torsion (Nieh-Yan topological current) and the Chern-Simons topological current which is thoroughly determined by the metric. Therefore, this quantum contribution has geometrical nature different from that one obtained in \cite{ptime}. We have also noted that this term resembles the Zeeman effect in the non-relativistic limit. 

In order to understand further the classical implications of the contact interaction term, we proposed a simple modified theory of gravity which consist of adding  the term (\ref{con}) to the EC action. In the absence of matter sources the field equations associated to the torsion tensor is non-trivial, though it remains non-dynamical as in EC case. Actually, we have seen that the torsion tensor is completely sourced by the metric as shown in Eq. (\ref{er}). As for the metric equation (\ref{37}), it imposes a constraint or consistency condition which leads to the vanishing of either $\tilde{C}_{\mu}$ or $\tilde{\nabla}_{\mu}\tilde{C}^{\mu}$.  
We have checked that the Schwarzschild solution persists in our modified theory, indeed the modified field equations reduces to the GR ones for all spherically symmetric metrics. On the other hand, Kerr metric cannot be a solution of this theory since it does not satisfy the consistency condition. As a non-trivial solution, we presented the G\"{o}del metric in which  the consistency condition is fulfilled, but the field equations for it do not reduce to GR ones.

It is interesting to note that our one-loop result is very similar to the known four-dimensional gravitational Chern-Simons term \cite{Jackiw}, with the role of the constant vector $b_{\mu}$ is played by the axial-vector torsion $S_{\mu}$. In principle, it allows to suggest that the Lorentz-breaking vectors in certain cases can be generated by non-zero expectation values (v.e.v) of the axial-vector torsion, so, as a by-product of our studies we arrived at a possible mechanism explaining the Lorentz symmetry breaking.

A natural continuation of this work would consist of exploring new solutions of this modified field theory. For example, a good candidate would be the G\"{o}del-type metrics and also finding a rotating black hole solution in this model since Kerr metric does not solve the modified equations of motion. Gravitational waves should be analyzed as well. We are examining these issues in a possible forthcoming work.

\textbf{Acknowledgments.}  
Authors are grateful to Gonzalo Olmo for helpful discussions and collaboration on related topics. The work by A. Yu. P. has been supported by the
CNPq project No. 301562/2019-9. P. J. Porf\'{i}rio would like to acknowledge the Brazilian agency CAPES for the financial support (PNPD/CAPES grant, process 88887.464556/2019-00), and also the Departament de F\'{i}sica Te\`{o}rica and IFIC, Universitat de Val\`{e}ncia, for the hospitality.

\end{document}